\documentclass[twoside]{ilcws08}
\usepackage[latin1]{inputenc}
\usepackage{wrapfig,rotating}
\usepackage{amssymb,amsmath,array}

\begin{document}

\title{
Realistic Simulation of the MAPS Response
}


\author{L.Maczewski$^1$, P.Niezurawski$^1$, A.F.\.Zarnecki$^1$, 
      M.Adamus$^2$ and J.Ciborowski$^1$
\thanks{
Supported from the research funds of the Polish Ministry of Science
and Higher Education as a part of the research project.} 
\vspace{.3cm}\\
%
1- Institute of Experimental Physics, University of Warsaw \\
Ho\.za 69, 00-681 Warszawa, Poland
\vspace{.1cm}\\
2- The Andrzej Soltan Institute for Nuclear Studies, \\
Ho\.za 69, 00-681 Warszawa, Poland
}


\maketitle

\begin{abstract}
The MAPS technology is considered as a possible choice for the ILC
Vertex Detector.  
Test results of MIMOSA-5 sensors indicate that the pixel multiplicity 
and the single point resolution depend significantly on the incident
particle angle. 
We propose a simple model describing charge distribution in the
detector, which can be used for detailed simulation of the Vertex
Detector response.
Good agreement with beam test data is obtained.
A new class for Track Detailed Simulation (TDS) has been developed and
implemented in the EUTelescope software framework.
\end{abstract}

\section{Introduction}

The optimisation of the design of the Vertex Detector (VXD)
for the experiment at the  International Linear
Collider (ILC) is based on
the Monte Carlo (MC) studies. In order to achieve a realistic
description of the VXD performances a detailed MC simulation is
required. 
Simulation studies performed so far focused mainly on the detector
geometry, assuming a given resolution of the pixel sensor.
However, dedicated studies show that the pixel multiplicity
and the single point resolution depend significantly on the angle of 
the incident particle.
Thus a Monte Carlo simulations should be supplemented by a 
digitisation procedure describing detector response to
charged particles on the pixel level. 
This was the motivation for developing a model of
signal formation for the Monolithic Active Pixel Sensor
(MAPS)~\cite{Deptuch1} which could be used in the full simulation of
the Vertex Detector. 

\section{Tests of the MAPS detectors}

\begin{figure}[!htbp] 
        \begin{center}
        \includegraphics[height=5.0cm]{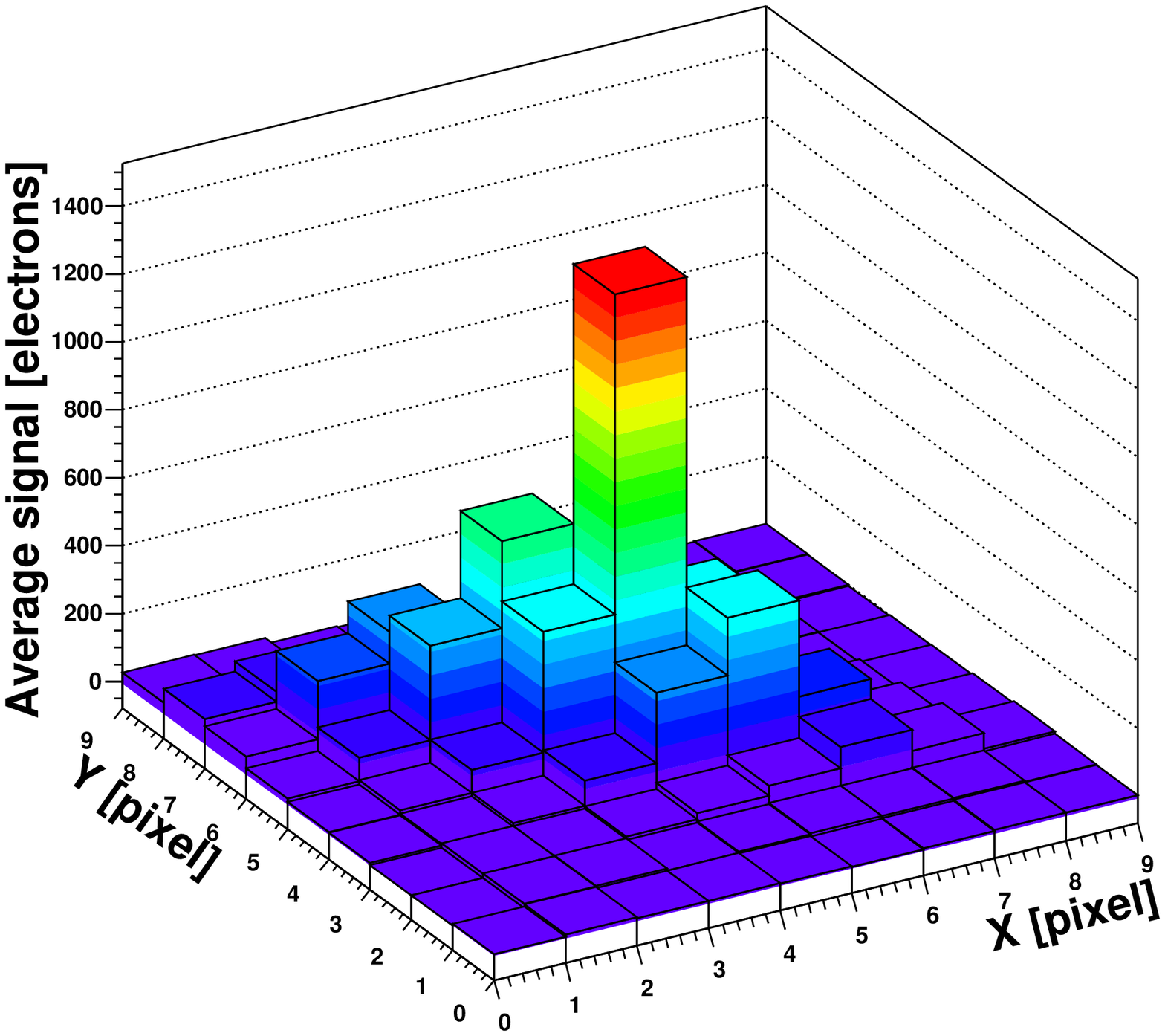}
        \hspace{1cm}
        \includegraphics[height=5.0cm]{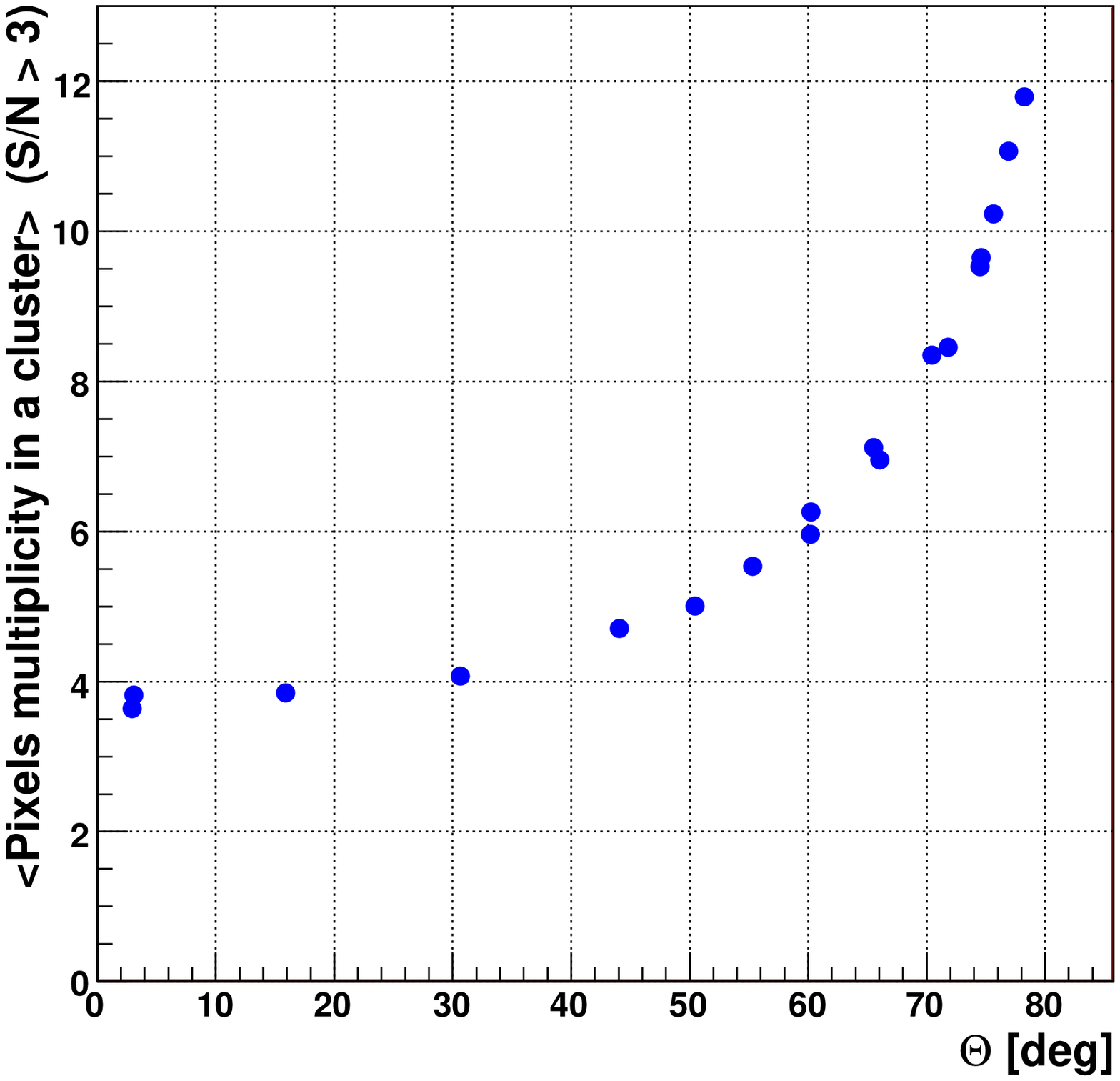}
        \end{center}
        \caption{Results of the MIMOSA-5 tests with 6~GeV electron
          beam: average cluster shape for beam incident angle
          $\theta = 75^{\circ}$ (left) and the mean pixel multiplicity
          in the cluster as a function of the track inclination
          $\theta$ (right).} 
        \label{testres}
\end{figure}

The MAPS technology provides a very good spatial resolution,
high signal to noise ratio, low material budget, low costs of
fabrication and high radiation tolerance. 
The Warsaw/Lodz group is involved in studies of the MAPS tracking
performances and signal formation. Two MAPS detectors have been
investigated: MIMOSA-5 with 17~$\mu$m pixel pitch and MIMOSA-18 with
10~$\mu$m pixel pitch, both with the epitaxial layer of 14~$\mu$m. 
The sensors have been exposed to the 6~GeV electron at DESY and
oriented at different angles w.r.t. the beam direction. 
An average cluster shape for MIMOSA-5, as measured for electrons incident
at  $\theta = 75^{\circ}$ is shown in fig.~\ref{testres}(left).
The elongation of the clusters is clearly visible. The dependence of
the number of pixels (with signal to noise ratio grater than three) on
the track inclination is shown in fig.~\ref{testres}(right). 

The increase of pixel multiplicity with the $\theta$ angle will influence the
VXD tracking performances and detector occupancy. This has to be taken
in to account in development of clustering algorithms and in the
single point resolution estimates. 

\section{Simple model of charge diffusion}

\begin{figure}[!b]
        \begin{center}
              \includegraphics[height=3.5cm]{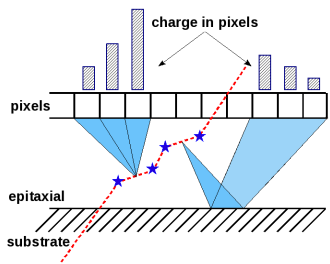}
                \caption{Schematic cross-section of the MAPS detector.}
                \label{SImpleModel}
        \end{center}
\end{figure}

There are three main layers of the MAPS device: a substrate, an
epitaxial layer and a pixel layer, see fig.~\ref{SImpleModel}.  
A charged particle traversing the detector ionise the silicon. 
In the proposed digitisation model for the MAPS detector three major
effects influence charge distribution. 
First, it is assumed that the charge generated in the epitaxial
layer diffuses isotropically. This means that charge carriers propagate
in all directions with the same probability. Approximately 50\%
of carriers move directly toward collecting diodes and the other 50\%
towards the substrate. We assume that after reaching the
substrate the carriers are reflected due to the potential barrier
present as an effect of a different doping concentration of the
substrate and the epitaxial layer, and that the reflection angle
is equal to the angle of incidence. 
Finally, we assume that the charge reaching the collecting diodes is
smaller that the primary ionization due to the trapping of the carriers
in the silicon. 
This simple model provides a good description of charge sharing in
the MAPS device for all incident angles, assuming the charge attenuation
length of about 50 $\mu m$.

\section{Implementation in the ILC Software}

Particles crossing at large angles can scatter inside the sensor
resulting in the increased ionisation path.
This is properly modelled in Geant. 
However, Mokka used to add all Geant steps in given layer and store
only single hit per particle, with total path length and energy deposit.
This made realistic simulation of the detector response impossible.
Therefore a dedicated option has been implemented in new Mokka release, which
forces Mokka to store separate Geant steps. It is currently implemented
for VXD only, but can be extended to other detectors if required.
For high energy particles from interaction point new option  increases
average number of VXD hits generated by Mokka by about 25\%.  
At the same time the number of hits with very large energy deposit
(above 10~mips) is suppressed by over an order of magnitude.

A new class for Track Detailed Simulation (TDS) has been developed
and implemented in the ILC software framework within the EUTelescope
package. 
For each Mokka hit, the charge deposit is divided into many steps
along particle path.
Charge collected in each pixel is then calculated by 
numerical integration of the charge diffusion formula.
Integration results are stored in a dedicated grid, so simulation is
very fast (except for first ~100 hits).
After the collected charge is calculated, gain, noise and analog to digital
conversion are taken into account.
Work is in progress to extend TDS package to the VXD detector simulation.

\begin{figure}[!t] 
        \begin{center}
                \includegraphics[height=4.0cm]{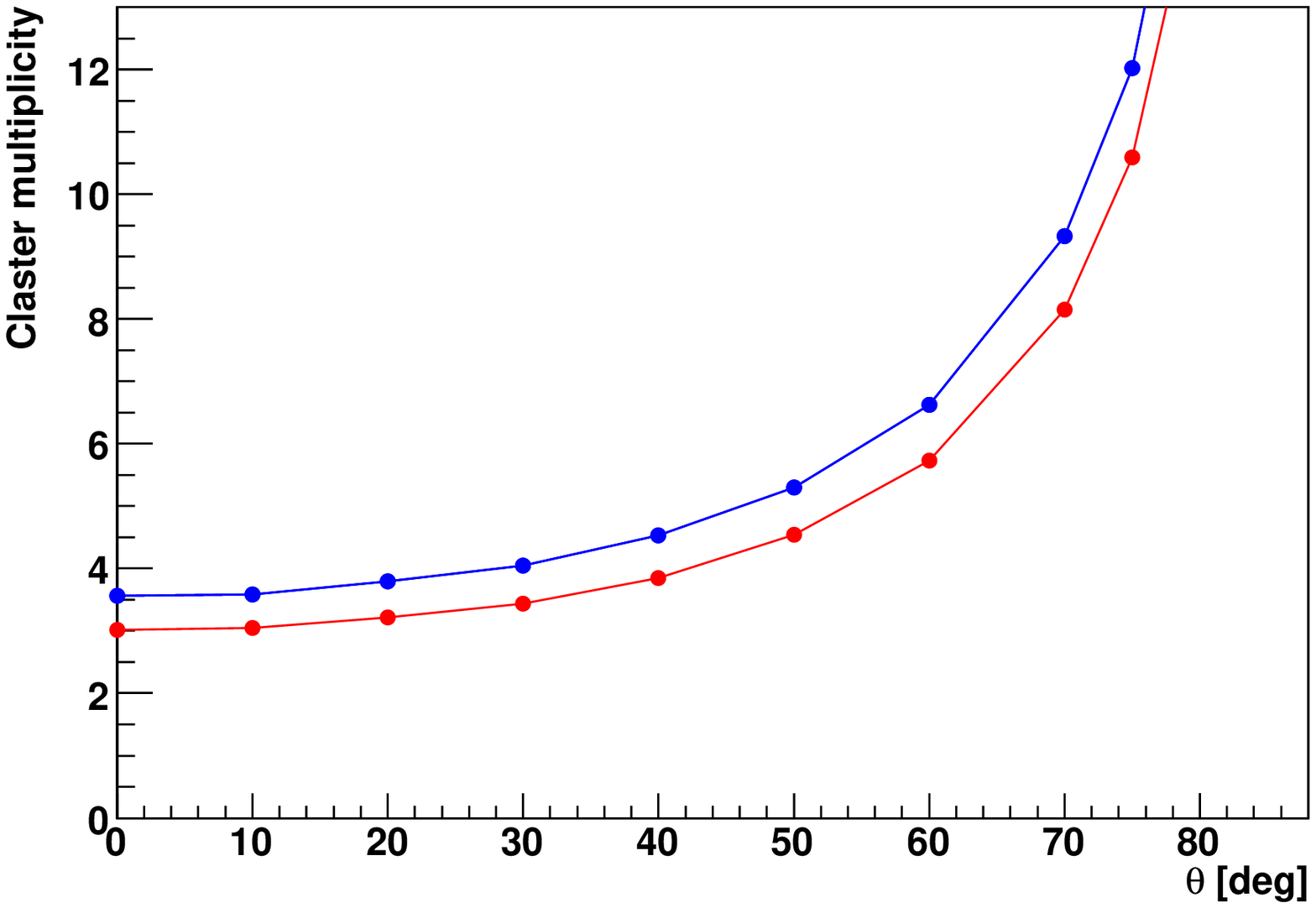}
        \hspace{0.5cm}
                \includegraphics[height=4.0cm]{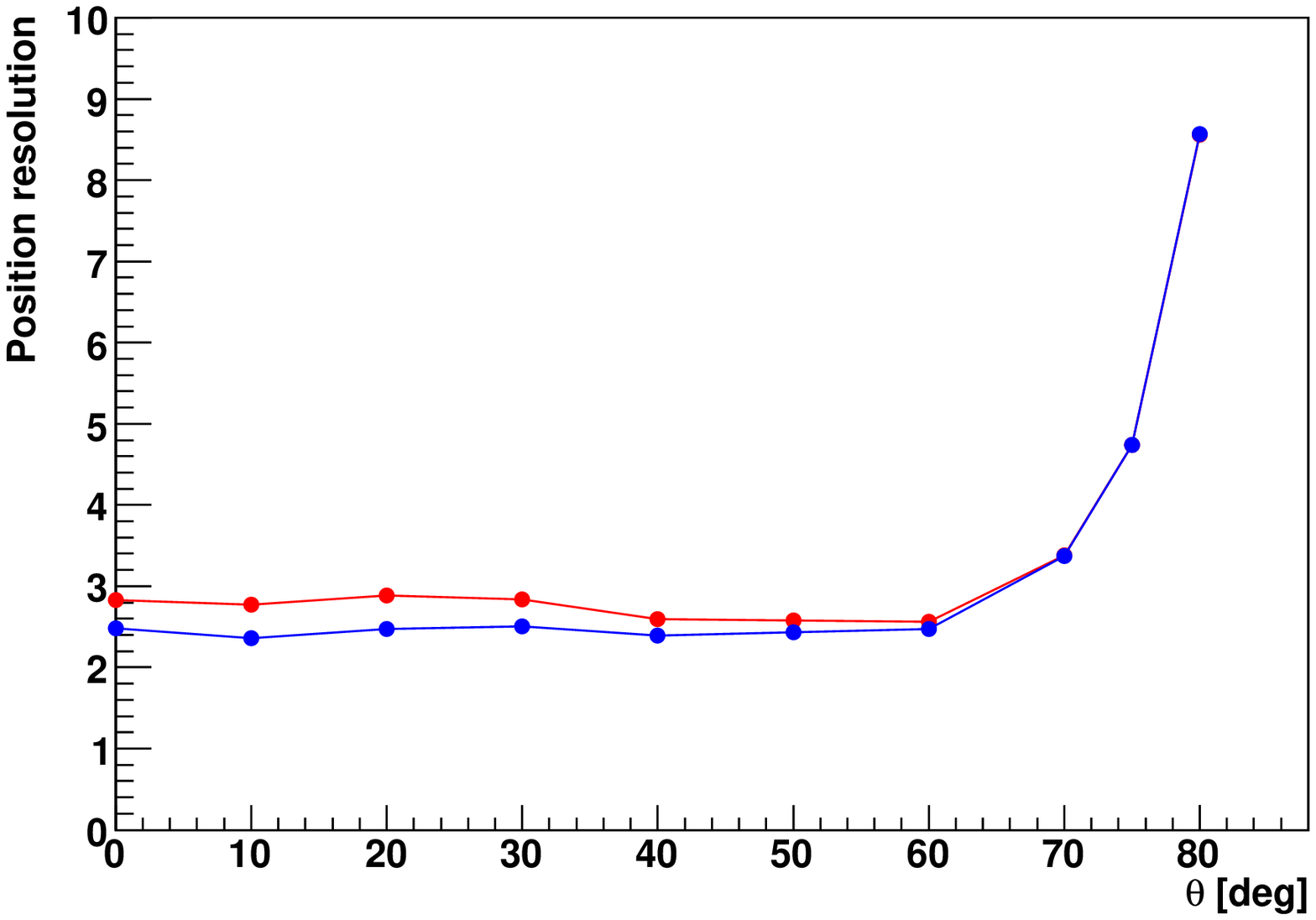}
        \caption{
First results of the MIMOSA-5 simulation with the TDS package:
mean cluster multiplicity for 6~GeV electrons (left) and the estimated position
resolution (right) as a function of the incident particle angle. 
} 
        \label{MC}
        \end{center}
\end{figure}

First, preliminary results of the MIMOSA-5 simulation with the TDS
package are presented in  fig.~\ref{MC}. 
Shown in the mean cluster multiplicity for 6~GeV electrons, as a
function of the incident particle angle, and the estimated position
resolution from center-of-gravity of 3x3 cluster around pixel with
maximum charge.  Two sets of results (indicated by blue and red lines) 
correspond to different digitization parameters assumed.


\begin{footnotesize}

\end{footnotesize}



\begin{thebibliography}{99}
\bibitem{url} Presentation: \\ 
\verb$http://ilcagenda.linearcollider.org/contributionDisplay.py?contribId=218&sessionId=21&confId=2628$

\bibitem{Deptuch1} G.~Deptuch {\it et~al.}, 
Nucl. Instr. and  Meth. Phys. Res. {\bf A511} 240 (2003). 

\end{thebibliography}
\end{document}